\documentclass{article}
\usepackage{spconf,amsmath,graphicx,cite}
\usepackage{amsmath,graphicx,amssymb,amsmath,epsfig,bm,float,multicol,multirow ,float,bm,rotating,array, url, hyperref}
\usepackage[table,xcdraw]{xcolor}

\newcommand{\ra}[1]{\renewcommand{\arraystretch}{#1}}
\title{On the compression of shallow non-causal ASR models using knowledge distillation and tied-and-reduced decoder for low-latency on-device speech recognition}
\name{Author(s) Name(s)\thanks{Thanks to XYZ agency for funding.}}
\address{Author Affiliation(s)}

\name{Nagaraj Adiga, Jinhwan Park, Chintigari Shiva Kumar,  Shatrughan Singh, Kyungmin Lee,  \hfill \break Chanwoo Kim, Dhananjaya Gowda}
\name{\parbox{\linewidth}{\centering Nagaraj Adiga, Jinhwan Park, Chintigari Shiva Kumar,  Shatrughan Singh, Kyungmin Lee,  Chanwoo Kim, Dhananjaya Gowda}}
\address{†Samsung Research}

\begin{document}
\ninept
\maketitle
\begin{abstract}
Recently, two-pass conformer-transducer architectures with a cascade of causal and non-causal encoders have emerged as a strong contender for on-device automatic speech recognition (ASR). 
These cascaded encoder models allow for decoding in both streaming and look-ahead modes while reducing the overall model footprint by sharing a common causal encoder as well as decoder.
One emerging trend to compress these cascaded encoder models further has been to fix a small budget size for the causal encoder, and compensating for any loss in accuracy by increasing the size of non-causal encoder.
On the contrary, in this paper, we propose to fix the budget size of the non-causal encoder and then use knowledge distillation to recover any loss in accuracy due to the reduction in the size of causal encoder. 
Further optimization is achieved by replacing the LSTM decoder with a a tied-and-reduced (TAR) decoder. 
The proposed TAR shallow non-causal conformer-T ASR model is compressed by reducing the causal encoder and TAR decoder layers proportional to the target compression before applying Kullback-Leibler divergence loss on the decoder posteriors against a teacher model.
We demonstrate a 50\% reduction in the size of a 41 M parameter baseline cascaded teacher model with no noticeable degradation in ASR accuracy and a 30\% reduction in latency.
\end{abstract}
\section{Introduction}
\label{sec:intro}
\vspace{-2mm}
Recent years have seen a flurry of end-to-end ASR architectures being proposed for cloud-based as well as on-device applications ~\cite{kim2020review,gulati2020conformer,garg2020streaming,variani2020hybrid,botros2021tied,dawalatabad2021two,rathod2022multi}.  
A few of these include connectionist temporal classification (CTC)~\cite{CTC}, listen attend and spell (LAS) network~\cite{LAS}, recurrent neural network transducer (RNN-T)~\cite{he2019streaming}, and Conformer Transducer (Conformer-T)~\cite{huang2020improving,gulati2020conformer}. 
Due to their low latency decoding capabilities conformer-T models have become a favourite for on-device applications. 
Two-pass models with a cascade of causal and non-causal encoders with two different decoders attached to each encoder for streaming and look-ahead decoding capabilities have been proposed in ~\cite{sainath2019two,huang2020improving,gulati2020conformer}. 
In ~\cite{narayanan2021cascaded}, a single shared decoder was proposed to train both the causal and non-causal encoders simultaneously, at the same time reducing overall model footprint. 

Reducing the memory footprints and latency of ASR models are crucial for on-device applications~\cite{he2019streaming}. 
There are multiple works proposed in the literature to compress ASR models and transducers models in particular. 
Knowledge distillation (KD) is one of the popular compression techniques used extensively in numerous works~\cite{panchapagesan2021efficient,dawalatabad2021two,rathod2022multi}. 
KD was applied to the larger transducer model in ~\cite{panchapagesan2021efficient}, and a reasonable compression was obtained. In~\cite{dawalatabad2021two} proved that KD also delivered a high compression rate with an RNN-T-based two-pass ASR model. 
In ~\cite{mirzadeh2020improved,rathod2022multi,garg2019improved}, the authors attempted to perform KD on a very large transducer model in two to three stages using progressive compression. 
In~\cite{variani2020hybrid,zhang2021tiny,zhang2020transformer}, the input to the LSTM of the decoder network is limited to two history phonemes without degrading WER. It is therefore proposed that, upon training, the LSTM be turned into a size-appropriate fast lookup table. 
Botros~\emph{et al.}~\cite{botros2021tied},  were then inspired to replace the LSTM layers of the transducer with a simple weighted averaging of the input embeddings and tie the embedding matrix weights to the joint network's output layer. This tied-and-reduced (TAR) technique significantly decreased the decoder's parameters without impacting the WER. This also suggests that encoder depth is more important for speech recognition performance than prediction network depth for transducer models.
However, relatively few studies have investigated the compression on cascaded encoder models.

One of the recent papers ~\cite{botros2023practical} attempts to reduce the causal encoder parameters to $\sim$50M of a large cascaded conformer-T model and further optimize it reduce the overall 1st pass latency.
In order to compensate for the loss in accuracy in 1st pass is compensated by increasing the 2nd pass non-causal encoder significantly under an high-latency application for 2nd pass decoding. 
However, in most practical ondevice applications the 2nd pass decoding also needs to have a low latency.
Motivated by this, in this paper we study multiple strategies for the compression of cascaded model while trying to reduce latency for both 1st as well as 2nd pass decoding.
In view of this, we propose to use a fixed-size (10M params) shallow non-causal encoder, replacing the LSTM decoder with a TAR decoder and then use KD to recover for any loss in accuracy due to reducing the size of the causal encoder.
The proposed TAR shallow-noncausal Conformer-T model is trained using KD from a larger baseline teacher model (41M params) to achieve more than 50\% compression without much degradation in WER. 
We additionally investigate optimizing the TAR decoder embedding layer dimension, encoder cell size, and the number of encoder layers of the shallow non-causal Conformer-T model. 
Furthermore, the shallow cascaded architecture reduces latency by 30\%.


\vspace{-3mm}

\section{Cascaded Model compression}
\label{sec:format}
\vspace{-2mm}
\subsection{Cascaded Conformer transducer}
\label{sec:conformerT}
Our ASR model is based on the cascaded Conformer-T architecture~\cite{narayanan2021cascaded}. The cascaded architecture consists of first pass causal encoder for generating immediate streaming output and second pass non-causal encoder to emit final output. Both encoders are made up of multiple layers of Conformer blocks~\cite{gulati2020conformer}. In addition, the cascaded model employs a shared transducer decoder that functions similarly to a language model. Both the causal and non-causal encoders directly connected to the shared decoder. During training, total loss is computed as weighted sum of output coming from shared decoder via causal and non-causal connections, respectively, as mentioned in~\cite{narayanan2021cascaded}. During inference time, model can operate either in streaming or non-streaming mode depending on feature extracted from causal or non-causal encoder for decoding. 

\vspace{-3mm}
\subsection{Tied-and-reduced Decoder}
\label{sec:TRP}
One method for performing model compression is using tied-and-reduced (TAR) decoder~\cite{botros2021tied}. 
The transducer network is altered for the TAR method, while the encoder part is kept the same as it was for the cascaded Conformer model. First, it uses embedding matrix to embed the previous time step label into embedding vector $V_i$. This embedding matrix is also tied with output layer of joint network. To keep the order of the last time step label position vector $S_i$ is added. The LSTM layers of the prediction network are then replaced by a simple weighted average of the input embeddings to have a reduced number of parameters. 
This prediction can be made with more than one head to have better performance~\cite{vaswani2017attention}. 
To make the model even better~\cite{botros2021tied}, the output of the prediction layer is projected, layer normalised~\cite{ba2016layer}, and then the swish nonlinear activation function~\cite{ramachandran2017searching} is applied. Overall, the changes from the TAR architecture led to a reduction of 90\% of Conformer-T decoder parameters. But this method only compresses the decoder network, so we need a way to compress both the encoder and decoder network to achieve a higher compression rate. 
\vspace{-3mm}
\subsection{Knowledge distillation Compression}
\label{sec:KD}
Knowledge distillation (KD) is utilized for compression by employing a teacher model and training a smaller student model to match the teacher model~\cite{hinton2015distilling}. In the initial stage, a larger teacher Conformer-T model is trained. In the second stage, the smaller student Conformer-T model is trained using the larger teacher model and distillation loss. In addition to the original training loss, the Kullback–Leibler (KL) divergence between the probability distributions of the teacher and student models as distillation loss is used to mimic the teacher model.

We followed the approach mentioned in ~\cite{panchapagesan2021efficient} for KD compression by specifying the student model using scaled-down encoder and decoder network parameters. Next, we train the compressed student model using KD from the teacher model. RNN-T loss and KL divergence loss are used to train the student model. The KL divergence between the output probability distributions of the teacher and student model can be given by:
$ L_{distill} = \sum_{u,t} \sum_{k} P^{T}(k|t,u)\ln \frac{P^{T}(k|t,u)}{P^{S}(k|t,u)}$ 
where $P^T$ and $P^S$ are output probability distributions of teacher and student Conformer-T models at time step $t$. The indexes $t$, $u$, and $k$ represent the input sequence length, output sequence length, and dimension of the output probability distribution, respectively. We can observe that distillation loss computation needs an additional memory size per utterance, which is too costly and impractical~\cite{panchapagesan2021efficient}. Therefore, in our implementation, we used the efficient distillation loss as described in ~\cite{panchapagesan2021efficient}. The total loss in KD compression training is determined by the linear combination of $L_{RNN-T}$ and $L_{distill}$ as follows: $L_{Total} = (1 - \alpha) L_{RNN-T} + \alpha L_{distill}$, where $\alpha$ denotes the empirically determined KD weight. 



\vspace{-3mm}
\subsection{TAR cascaded Conformer-T compression}
\label{sec:KD+TRP}
\begin{figure}[t]
\centerline{\epsfig{figure=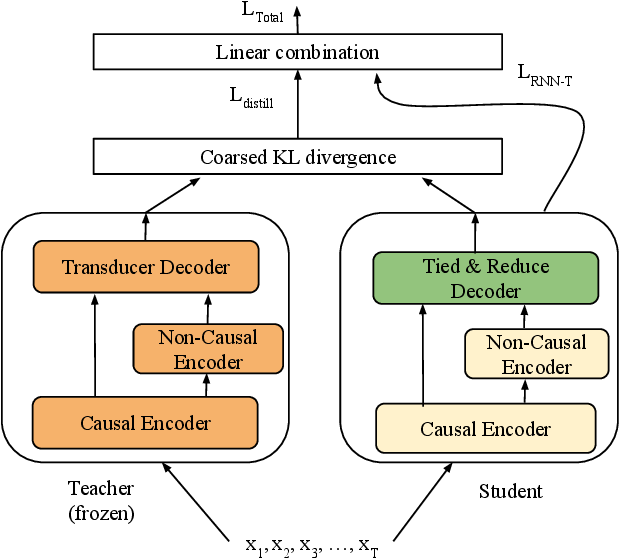,width=55mm}}
\vspace{-2mm}
\caption{The block diagram of suggested shallow cascaded model using TAR and KD method. During student training, orange-colored blocks are frozen, and the LSTM-based transducer network is replaced with a TAR network. The student model was distilled from the teacher model using KL divergence Loss.}
\label{img:block_diagram}
\vspace{-5mm}
\end{figure}

\vspace{-1mm}
To achieve a high compression rate, and inspired by the TAR technique~\cite{botros2021tied}, we merged TAR architecture with KD compression in our proposed approach. We used a TAR network in place of the LSTM decoder. A scaled-down cascaded Conformer encoder was also used while defining the student model. Therefore, in the suggested improvements, we directly applied KD to the TAR network rather than the reduced LSTM decoder. Furthermore, we see that the increased latency of the cascaded model is primarily attributable to the addition of a non-causal encoder. As a result, we also compressed the non-causal encoder, which we will now refer to as a shallow cascaded model. Fig.~\ref{img:block_diagram} shows a summary of the changes that have been proposed. We first train a standard, larger cascaded Conformer-T model using RNN-T and L2 loss~\cite{narayanan2021cascaded,gulati2020conformer}. We freeze the teacher model and use a TAR  network instead of the LSTM transducer model for a training student model. Then, as shown in the subsection~\ref{sec:TRP}, training of the shallow model is carried out using KL divergence and RNN-T loss. The rationale behind this method is that since the decoder network does not require a longer history context, relatively the decoder network parameter can be reduced better with TAR approach rather than directly scaling down number of parameters. To train the student model, we also used a teacher model with a TAR cascaded Conformer-T model configuration. Unfortunately, because the RNN-T alignment was not converged, the student model did not train well.
\vspace{-3mm}

\section{Experimental Setup}
\label{sec:exp}
\vspace{-2mm}
We used the Librispeech database for our compression studies, which has 960 hours of training data. The "test-clean" (5.4 hours includes 2620 sentences) and "test-other" (5.1 hours includes 2939 sentences) sets are used to evaluate our models. As input to the cascaded Conformer model, we used 80 Mel-filter bank features extracted for every 10 ms computed over a window of 25 ms. The target vocabulary contains 1k byte pair encoding (BPE) units~\cite{sennrich2015neural,singh2021comparative}. For training, we applied a SpecAugument-based data augmentation technique~\cite{park2019specaugment}. A beam size of 4 is used for decoding. All our experiments are conducted on a two A100 GPU with an average batch size of 128, due to paucity of resources. We trained all models with mixed precision~\cite{micikevicius2017mixed} to achieve a larger batch size. This allows us to maintain nearly twice the batch size without compromising overall accuracy. In our experiments, the objective metrics we employed were the word error rate (WER).
\vspace{-3mm}
\subsection{Cascaded Conformer-T model}
\vspace{-1mm}
As we discussed in section~\ref{sec:conformerT}, the cascaded Conformer-T model is made up of two encoders (causal and non-causal) and shared decoder network. We used medium-sized Conformer-T model~\cite{gulati2020conformer} . The causal and non-causal network consists of 16 and 6 layers of Conformer blocks. The decoder network is made of 2 layers of LSTM. The joint network has a feed-forward layer. Tensorflow 2 is used to build the whole training pipeline. We use a learning rate of 0.0003 and slow it down using the decay function described in ~\cite{rathod2022multi}. In our experiments, we used a dropout of 0.1 and causal weight of 0.8.
\vspace{-3mm}
\subsection{TAR cascaded Conformer-T}
\vspace{-1mm}
We tied the input embedding matrix to the joint network in the TAR network. Additionally, to reduce the total parameter, the dimension of the joint network's and the embedding layer's last layer was maintained the same~\cite{botros2021tied}. The embedding matrix is also shared over the history of the previous output token to obtain the embedding vector. In our experiments, we employed a history token of size 5. To increase the model's capacity and improve prediction, we additionally used a multi-head of 4. We chose these values based on empirical evidence. We can further reduce TAR parameters by scaling down the size of the embedding layer.
\vspace{-3mm}
\subsection{Knowledge distillation on shallow cascaded model}
\vspace{-1mm}
The non-causal encoder uses beam search for decoding, which causes an increase in second pass decoding latency. Therefore, we developed shallow cascaded model by reducing the size of non-causal encoder and applied KD on non-causal encoder.
Two different KD models are trained. One involves using KD on the TAR cascaded Conformer-T model. The second involves applying KD to a regular cascaded Conformer-T model, which serves as the compression baseline for our experiments. We scaled down the Conformer model by decreasing hidden units and/or layers of the both causal/non-causal encoder network. The model parameters of each of the employed teacher and student models can be found in Table~\ref{table:encoder_decoder_compression}. Other attributes remain the same across all models. We trained each model for 200 epochs and then published the results for the epoch that produced the best WER. According to our initial studies, the optimum outcomes for knowledge distillation come from a distillation loss weight of 0.02. The temperature was set to 1.0 for all our teacher and student model experiments.
\vspace{-3mm}

\section{Results and Discussion}
\label{sec:typestyle}
\vspace{-1mm}
This section demonstrates the effectiveness of our proposed compression technique for the TAR cascaded Conformer-T model. The various compression factors are achieved by making the encoder and decoder's cell size smaller and/or reducing the number of layers. As explained in Section~\ref{sec:TRP}, we tried different compression rates and evaluated how well they worked. In these experiments, we evaluated model for both streaming (S) and  non-streaming (NS) case. In addition, no results utilized a language model.
\vspace{-4mm}
\subsection{Performance of proposed KD Compression on TAR cascaded Conformer-T model }
\begin{table}[h]
\caption{Experimental results using KD on cascaded conformer-T (E1) and TAR conformer-T models (E2). The WER results are reported in \% for non-streaming (NS) and streaming (S) mode.}
\label{table:encoder_decoder_compression}
\centerline{
\ra{1}
\scalebox{0.7}{
\begin{tabular}{|c|c|c|c|c|c|}
 \hline
 \multicolumn{1}{|c|}{\bf{Comp.}} & \multicolumn{1}{c|}{\bf{\# Model}} & \multicolumn{2}{c|}{\bf{test-clean }} & \multicolumn{2}{c|}{\bf{test-other }}\\
 \cline{3-4}\cline{5-6}
 {\bf Factor} &  {\bf params (M)  } & {\bf NS} &  {\bf S } & {\bf NS } &  {\bf S }\\
\hline  
\multicolumn{6}{|c|}{\bf E0:  Baseline cascaded Conformer-T models}\\
\hline  
-  & 41.0 & 4.60  & 6.04 & 13.01 & 16.10\\
40  & 24.5 & 4.91  & 6.92 & 14.15 & 17.50\\
50  & 20.0 & 5.16  & 7.23 & 14.71 & 18.73\\
\hline
\multicolumn{6}{|c|}{\bf E1:  Compression  on cascaded Conformer-T model}\\
\hline  
40 & 24.5& 4.59 & 6.34 & 13.72& 16.84 \\
50 & 20.5 & 4.78 & 6.51 & 14.24 & 18.01 \\
60 & 16.0 & 5.96 & 9.02 & 16.15& 22.13 \\
70 & 12.6 & 7.04 & 11.24 & 19.16 & 26.35 \\
\hline
\multicolumn{6}{|c|}{\bf{\noindent E2: Compression on TAR cascaded Conformer-T model}}\\
\hline
40 & 24.0	& 4.47	& 6.20	&13.24	&16.49\\
50 & 20.0	& 4.62	& 6.32	&13.66	&17.16\\
60 & 16.5	& 4.83	& 8.49	&14.68	&20.63\\
70 & 12.5	& 6.13	& 11.04	&16.54	&25.61\\

\hline

\end{tabular}}}
\vspace{-3mm}
\end{table}
First, we conducted baseline experiments (E0) and trained three cascaded Conformer-T models. A larger model with 41 M parameters serves as the teacher for all student models during training. In addition, we trained two smaller models with 24.5 M and 20 M parameters to examine how well the smaller models performed without the use of any KD based compression techniques. The baseline results of all three models are reported in Table~\ref{table:encoder_decoder_compression}.
Our baseline streaming conformer results (6.04/16.10) are closer but poorer to a recent conformer baseline (5.56/13.23) reported in ~\cite{conmer-is2023}, which could be attributed to a smaller effective batch size in our case with only 2 A100 GPUs as against 8 in ~\cite{conmer-is2023}.
Our baseline non-causal or non-streaming performance (4.60/13.01) should be seen as a clear improvement over the 1st pass streaming accuracy.
It cannot be compared against the fully non-causal conformer accuracy (3.72/8.81) reported in ~\cite{conmer-is2023}, as our model uses the same shared causal encoder from the 1st pass during the 2nd pass non-streaming decoding.

In the first experiment (E1), we made a baseline compression model in which the size of the cascaded Conformer-T encoder and decoder networks was directly scaled down. The smaller student model is trained by distilling the knowledge from the teacher using KL divergence loss as described in section~\ref{sec:KD}. The results of KD compression are shown in Table~\ref{table:encoder_decoder_compression} for compression factor range 40-70\%. The table shows that we achieved a compression factor of 50\% with less than 5\% relative word error rate (RWER) degradation compared to the baseline in the case of the non-streaming mode for both "test-clean" and "test-other" test sets. However, for streaming mode RWER degradation is more than 5\% for both test sets at 50\% compression level. From the table, we can also see that compression with the aid of KD resulted in a reduction of at least 0.3 absolute WER when compared to directly training smaller models (E0: baseline experiments). This also shows the significance of applying KD for compressing shallow cascaded models.

In our second experiment (E2), we applied compression to the proposed TAR cascaded Conformer-T model. As explained in section~\ref{sec:KD+TRP}, we reduced the size of the transducer network by employing the TAR architecture and direct compression of the encoder network. The results of the proposed method for "test-clean" and "test-other" test sets are shown in Table~\ref{table:encoder_decoder_compression}. From the table, we can see that for the proposed compression, we achieved a compression rate of 50\% in the case of the both "test-clean" and "test-other" test cases with less than 5\% relative WER. Additionally, in the case of the "test-clean" test set with non-streaming mode, we achieved a compression rate of 60\% with less than 5\% RWER reduction. We can see that the TAR cascaded Conformer-T compression method performs better than the cascaded Conformer-T model at each compression rate. These findings demonstrate the usefulness of combining TAR and KD compression approaches to reduce model footprint. At a lower level of compression rate (40\% compression in Table), we also notice a drop in WER in comparison to baseline. This can be as a result of the additional teacher model that is available to guide the student model. This behavior is even noted in literature~\cite{panchapagesan2021efficient,rathod2022multi}.
\vspace{-3mm}
\subsubsection{Latency evaluation on the mobile system}
The latency evaluation (experiment E3) of the proposed shallow cascaded encoder is shown in Table~\ref{table:latency_evaluation}. This latency value is the time taken to process the complete audio after the end of utterance occurs. 
This measurement includes the entire operations for ASR including feature extraction, first pass causal encoding and second-pass beam-search decoding. The evaluation reported in the table computed for 1k test utterances. We saw a 30\% reduction in latency time (89 ms) for 50\% compressed model compared to the uncompressed baseline model (128 ms) computed on the snapdragon SM8450 processor. Here, we only included latency evaluation up to a 50\% compressed model since, beyond that, even though latency values are dropping, the overall RWER degradation is still quite large in comparison to baseline models. The proposed system's real-time factors (xRT) are also showed in ~\ref{table:latency_evaluation}.  These findings are consistent with the latency analysis.
\begin{table} [t]
\caption{E3: Latency and real time factor (xRT) measured on Sanpdragon SM8450 processor of baseline and proposed shallow cascaded model}
\label{table:latency_evaluation}
\centerline{
\ra{1}
\scalebox{0.7}{
\begin{tabular}{|c|c|c|c|c|}
  \hline
 \multicolumn{1}{|c|}{\bf{Comp.}} & \multicolumn{1}{|c|}{\bf{Model}} &
   \multicolumn{1}{c|}{\bf{Metric}}& \multicolumn{2}{c|}{\bf{Snapdragon SM8450}} \\
   
  \cline{4-5}
 {\bf factor (\%)} & {\bf Params (M)} &  {\bf }  & {\bf 1st Pass} &  {\bf 2nd Pass } \\
\hline  \hline
Baseline & 41.0 & latency (ms) & 48.2  & 128 \\
         &            & xRT & 0.14  & 0.15\\
\hline  \hline
40 & 24.5 & latency (ms) & 42.1  & 98.8 \\
   &       & xRT & 0.10  & 0.12\\
          
\hline  \hline
50 & 20.0 & latency (ms) & 36.4  & 89.0 \\
   &       & xRT & 0.08  & 0.09\\
\hline
\end{tabular}}}
\vspace{-3mm}
\end{table}
~\vspace{-1mm}

\vspace{-3mm}
\subsection{Ablation studies}
We used the best mix of compression on the encoder and decoder networks in the proposed shallow cascaded architecture. To determine the efficacy of each module, we independently compressed the embedding layer dimension, encoder cell size, and encoder layers. This is described in more detail in the section below.

\vspace{-3mm}
\subsubsection{Effect of decoder only compression using TAR}
\vspace{-1mm}
In experiment 4 (E4), we intended to figure out the TAR cascaded Conformer-T model's adequate embedding size. As mentioned in Section~\ref{sec:TRP}, we employed a similar tied embedding matrix and reduced the prediction network by averaging input embeddings. We kept the embedding size at 768 in the baseline Conformer-T model. Then, we changed the input dimension of the embedding network by a factor of 1x, 0.5x, and 0.25x, respectively, compared to the baseline. Table~\ref{table:embedding_size} displays this experiment's findings with varying the embedding dimension's size. The values in the table's second column indicate how many total parameters there are for decoder network. Based on the results, it is evident that we can reduce the decoder network parameters by as much as 95 \% without a significant increase in the overall WER. 
This also align with results reported in~\cite{botros2021tied} that size of encoder network is more important than that of decoder network.
On the contrary, the WER improves significantly for streaming decoding implying that the TAR decoder need not be fixed at 1x embedding size can be allowed to be compressed as is done in our E2 experiments.
Embedding size smaller than 0.25x saw an increase in overall WER.

\begin{table} [t]
\caption{E4: Decoder compression using only TAR network}
\label{table:embedding_size}
\centerline{
\ra{1}
\scalebox{0.7}{
\begin{tabular}{|c|c|c|c|c|c|c|}
 \hline
 \multicolumn{1}{|c|}{\bf{\# Embed.}} &
  \multicolumn{1}{c|}{\bf{\# Model}} & \multicolumn{1}{c|}{\bf{Comp.}} & \multicolumn{2}{c|}{\bf{test-clean}} & \multicolumn{2}{c|}{\bf{test-other}}\\
  \cline{4-5}\cline{6-7}
 {\bf size} &  {\bf params} & {\bf \%  } & {\bf NS} &  {\bf S } & {\bf NS } &  {\bf S }\\
\hline  \hline
Baseline & 7.00 & - & 4.60  & 6.04 & 13.00 & 16.10\\
768 (1 x) & 1.60 & 77 & 4.37  & 6.00 & 12.99 & 16.69\\
384 (0.5 x) & 0.65 & 90 & 4.46  & 6.04 & 12.72 & 16.15 \\
192 (0.25 x) & 0.29 & 95 & 4.22  & 5.82 & 12.72 & 16.01\\
\hline
\end{tabular}}}
\end{table}

\vspace{-3mm}
\subsubsection{Effect of encoder cell size for KD compression}
\vspace{-1mm}
Next, we want to know how much the encoder model of the Conformer-T architecture can be compressed. Consequently, in experiment 5, we compressed the encoder using the knowledge distillation technique described in section~\ref{sec:KD} while keeping the settings of the decoder network the same as that of the baseline. In encoder-only compression, we preserved the same number of layers as in baseline and just altered the quantity of encoder-hidden units with the compression factor as displayed in the first column of Table~\ref{table:encoder_cell_size}. Only the causal encoder network part was compressed in this investigation, as it is shared by both streaming and non-streaming decoding modes. Based on the results, it is evident that we can compress the encoder network hidden units by up to 50\% sparsity without significantly impacting the overall WER.
\begin{table} [t]
\caption{E5: Encoder cell-size compression using KD on cascaded Conformer-T}
\label{table:encoder_cell_size}
\centerline{
\ra{1}
\scalebox{0.7}{
\begin{tabular}{|c|c|c|c|c|c|}
  \hline
 \multicolumn{1}{|c|}{\bf{Comp.}} &
   \multicolumn{1}{c|}{\bf{\# Model}}& \multicolumn{2}{c|}{\bf{test-clean}} & \multicolumn{2}{c|}{\bf{test-other}}\\
  \cline{3-4}\cline{5-6}
 {\bf factor (\%)} &  {\bf params (M)}  & {\bf NS} &  {\bf S } & {\bf NS } &  {\bf S }\\
\hline  \hline
Baseline & 41  & 4.60  & 6.04 & 13.00 & 16.10\\
40          & 25 & 4.57  & 6.12 & 13.29 & 16.54\\
50          & 20 & 4.59  & 6.35 & 13.42 & 16.98\\
58          & 17 & 5.01  & 7.60 & 14.15 & 19.23\\
65          & 14 & 5.50  & 9.41 & 15.66 & 22.71\\

\hline
\end{tabular}}}
\vspace{-3mm}
\end{table}

\vspace{-3mm}
\subsubsection{Effect of reducing number of encoder layers}
\begin{table} [t]
\caption{E6: Encoder layer-depth compression of TAR cascaded Conformer-T}
\label{table:encoder_layer}
\centerline{
\ra{1}
\scalebox{0.75}{
\begin{tabular}{|c|c|c|c|c|c|c|}
 \hline
   \multicolumn{1}{|c|}{\bf{\# Encoder}} &
 \multicolumn{1}{c|}{\bf{\# Model}}&
  \multicolumn{1}{c|}{\bf{Comp.}} &
    \multicolumn{2}{c|}{\bf{test-clean}} & \multicolumn{2}{c|}{\bf{test-other}}\\

  \cline{4-5}\cline{6-7}
 {\bf Layers} &  {\bf params} & {\bf factor (\%)  } & {\bf NS} &  {\bf S } & {\bf NS } &  {\bf S }\\
\hline  \hline
Baseline & 41.0 & - & 4.6  & 6.04 & 13.0 & 16.10\\
\hline
\multicolumn{7}{|c|}{\bf{Compression on TAR Conformer-T model}}\\
\hline
16 & 20.0 & 51.6 & 4.62  & 6.32 & 13.66 & 17.16\\
15 & 19.3 & 53.0 & 5.26 & 8.01 & 15.22 & 20.08 \\
14 & 18.8 & 54.5 & 5.41  & 8.24 & 15.34 & 20.25\\
13 & 18.2 & 56.0 & 5.52 & 8.91 & 15.53 & 21.32\\
\hline
\end{tabular}}}

\end{table}
\vspace{-1mm}
In this experiment six (E6), we fixed the optimal hyper-parameters we obtained from prior ablation investigations with TAR network embedding size as 0.25x and encoder network cell size compressed to 50\%. We cut down on the number of encoder network layers to see if we could make the TAR cascaded Conformer-T model even smaller. In both ``test-clean" and ``test-other" test cases, relative word error rate is more than 10\% for non-streaming mode as shown by Table~\ref{table:encoder_layer}. Further, we can notice that the word error rate has significantly increased for streaming mode (RWER $>$ 20\%). This implies that deeper layers are crucial for encoder modelling.
\vspace{-3mm}
\section{Conclusions}
\vspace{-1mm}
In this paper, we explore multiple strategies for compression of a cascaded Conformer-Transducer model. We investigated model compression techniques such as knowledge distillation, shared decoder, Tied-And-
Reduced (TAR) networks, and their combinations. We initially replaced the shared decoder network of a cascaded Conformer-T model with TAR network. Additionally, with the aid of larger teacher model TAR cascaded Conformer-T model is further compressed using knowledge distillation technique. The proposed method begins with a 41 M parameter streaming/non-streaming teacher model and reduces it to a shallow cascaded model without impacting the speech recognition accuracy. With the recommended {\it shallow non-causal architecture}, we can obtain the model up to 20 M parameters with a 50\% compression. We observe that the relative WERs for the suggested model are 4\% and 2\%, respectively, on the test-clean and test-other LibriSpeech datasets. Furthermore, the proposed shallow cascaded model reduces the latency time by 30\%. 
\vfill

\bibliographystyle{IEEEbib}
\bibliography{refs}
\end{document}